\documentclass{sf2a-conf2011}
\pdfoutput=1
\usepackage{graphicx}
\usepackage{hyperref}
\usepackage[]{natbib}  
\usepackage[cyr]{aeguill}
\usepackage{epstopdf}

\def\BibTeX{{\rm B\kern-.05em{\sc i\kern-.025em b}\kern-.08em
    T\kern-.1667em\lower.7ex\hbox{E}\kern-.125emX}}
\bibpunct{(}{)}{;}{a}{}{,}  

\begin{document}

\TitreGlobal{SF2A 2011}


\title{Calibrating 15 years of GOLF data}

\runningtitle{GOLF 15 years}

\author{G. R. Davies}\address{Laboratoire AIM,
  CEA/DSM-CNRS-Universit\'{e} Paris Diderot; IRFU/SAp, Centre de
  Saclay, 91191 Gif-sur-Yvette Cedex, France}

\author{R. A. Garc\'{i}a$^{1}$}



\setcounter{page}{237}

\index{Davies, G. R.}
\index{Garc\'{i}a, R. A.}


\maketitle


\begin{abstract}
The GOLF resonant scattering spectrophotometer aboard SoHO has now provided 15 years of continuous high precision Sun-as-a-star radial-velocity measurements.  This length of time series provides very high resolution in the frequency domain and is combined with very good long-term instrumental stability.  These are the requirements for measuring the low-l low-frequency global oscillations of the Sun that will unlock the secrets of the solar core.  However, before the scientifically interesting gravity and mixed modes of oscillation fully reveal themselves, a correction and calibration of the whole data set is required.  Here we present work towards producing a 15 year GOLF data set corrected for instrumental ageing and thermal variation.
\end{abstract}



\section{Introduction}

The Global Oscillations at Low Frequency (GOLF) instrument measures
the Sun-as-a-star radial velocity field \citep{GabGre1995,1997SoPh..175..207G}.  Raw measurements are made in
intensity and then corrected and calibrated to produce final residual
velocities \citep{2005A&A...442..385G}.  The ability of this calibration process to
produce high quality and very long time series (15 years) is important
as GOLF is forced to operate in a single wing mode \citep{1999A&A...341..625P}.  The
scientific motivation for an improved calibration comes from the need
for a stable 15 year time series.  This will allow detailed analysis
of the impact of the solar cycle on helioseismic parameters \citep{2009A&A...503..241B,2009A&A...504L...1S} and the
detection of new low-frequency oscillations.  Low-frequency
oscillations have low signal-to-noise ratios making their detection a
significant challenge.  Recent reports on the signature of g modes
\citep{2004ApJ...604..455T} and their fingertips \citep{2007Sci...316.1591G,2008AN....329..476G} sparked much debate.  An improvement
in GOLF performance at low frequency could reveal the secrets of the
solar core.
  
\section{GOLF observations}

\begin{figure}[ht!]
 \centering
 \includegraphics[width=0.8\textwidth,clip]{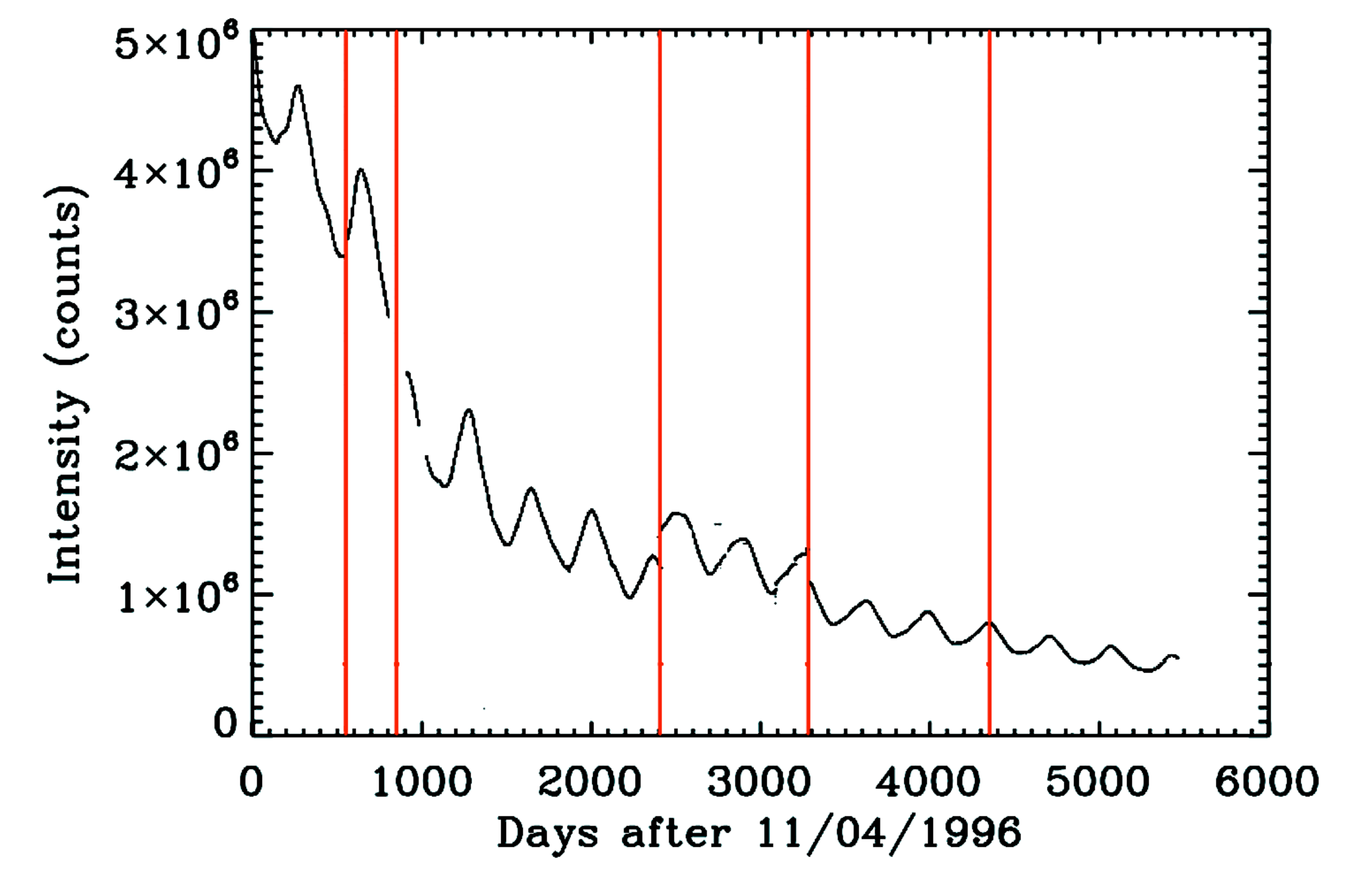}      
  \caption{GOLF raw-intensity measurements as a function of time.  Red
  lines mark breaks in continuity due to instrumental adjustments.  The
obvious effects displayed are the ageing of the photomultiplier tubes
(PMT near exponential decay), the Sun-instrument line-of-sight
velocity (one year period), and the discontinuities due switching
between the ``red'' and ``blue'' wings and PMT high voltage adjustment.}
  \label{davies:fig1}
\end{figure}

GOLF makes measurements of intensity, $I$, integrated over narrow
bands in wavelength, $\lambda$, and integrated over the visible
surface of the solar disk, $S$.  We can state this as a simple model,
\begin{equation}
I = \int_{\lambda} \int_{S}I_{\odot}(\lambda, S, v_{\rm los})
W(\lambda, S, v_{\rm los}) \; {\rm d}\lambda \; {\rm d}S ,
\end{equation}
where $I_{\odot}$ is the solar D1 and D2 absorption line spectrum, $W$
is the instrumental weighting, and $v_{\rm los}$ is the line-of-sight
velocity between GOLF and the Solar surface.  Given the measured
intensity, the solar line function, and the instrumental weighting it
would be possible to determine $v_{\rm los}$.  However, all three
components contain some level of uncertainty and/or cannot be
sufficiently modeled \citep{2000A&A...364..799U}.  So we make simplifications and propose a
method that relies on the following statement: a well corrected
instrument will give intensity measurements that are only a function
of $v_{\rm los}$.  This gives us the definition of a parameter for
good - the correspondence of intensity and line-of-sight velocity.\\
To achieve the best possible GOLF signal-to-noise ratio, it is
necessary to correct for long term instrumental effects and to provide
a stable calibration that minimizes non-solar variation in mode
amplitude.  Previous work has tackled these two aspects one after
another.  However, the gain of the instrument (photons detected
divided by photons collected) and the instrument sensitivity (${\rm
  d}I / {\rm d}v_{\rm los}$) are coupled.  The two processes must be
treated concurrently to give an optimum calibration.  Figure \ref{davies:fig1} shows
GOLF raw-intensity measurements as a function of time.  There are two
processes that are clearly visible: the yearly variation in the
Sun-instrument line-of-sight velocity; and the ageing of the
instrument.  Further to this there is a less obvious variation of
instrumental gain which is a function of the temperature of the
instrument.\\
To account for all effects present in Fig. 1 we modify equation 1.  We
add the ageing of the instrument (probably dominated by the ageing of
the photomultiplier tubes) $P(t)$, the variation in flux as a
function of the spacecraft-Sun distance $F(d)$, and the variation in
gain due to the changes in instrument temperature $K(T)$.  This gives
\begin{equation}
I = P(t) F(d) K(T) \int_{\lambda} \int_{S}I_{\odot}(\lambda, S, v_{\rm los})
W(\lambda, S, v_{\rm los}) \; {\rm d}\lambda \; {\rm d}S.
\end{equation} 
Given the simple model of the the underlying system and a
goodness-of-fit parameter, we can use the 15 years of GOLF observation
to determine instrumental parameters.  First we must define the
parameters and functions describing the instrument.\\
The ageing of the instrument is expected to de dominated by the
exponential loss of efficiency in GOLF's photomultiplier tubes.  From
GOLF's measurements we find that the rate of ageing varies as a
function of time.  To describe this in the simplest possible manner we
use the following function,
\begin{equation}
P(t) = p_{0} \exp{ -(t(p_{1} + p_{2}t))}.
\end{equation}
The change in flux due to the variation in the spacecraft-Sun distance
is determined by the solid angle presented by the GOLF aperture.  It
is trivial to how that 
\begin{equation}
F(d) = f_{0}^{2} / d^{2},
\end{equation}
where $f_{0}$ is a reference distance.\\
For the gain of the instrument due to temperature we use the form of
\cite{2005A&A...442..385G},
\begin{equation}
K(t) = 1 - k_{1} (T - k_{0}) - k_{2} (T - k_{0})^{2},
\end{equation}
where $k_{0}$ is a reference temperature.  Variations in
instrumental gain due to temperature have a number of possible sources:
variation in the vapour cell stem temperature; variation in the
detector temperature (PMT and counting electronics); and other sources
including the temperature of the interference filters.  Rather than
considering a function for each temperature we note that each of
the measured temperatures (particularly the cell stem and detectors) are
well correlated.  Using this we apply a single correction using only the
stem temperature, which is the measured temperature with the greatest
precision.\\  
We must also describe the sensitivity of the instrument (${\rm d}I/{\rm
d}v_{\rm los}$).
In this work we use the derivative of a fourth order
polynomial as we know this has been successfully applied to similar
but ground based instruments \citep{1995A&AS..113..379E,2009MNRAS.396L.100B},  
\begin{equation}
\frac{{\rm d}I}{{\rm d}v_{\rm los}} = \sum _{i=1}^{4} a_{i} \; i \; v_{\rm los}^{i-1}. 
\end{equation}
We have developed more sophisticated models of the instrumental
sensitivity that require more space for description than is available
here.  The full description will be presented in Davies et al. (in prep.).

\section{Results}

\begin{figure}[ht!]
 \centering
 \includegraphics[width=0.8\textwidth,clip]{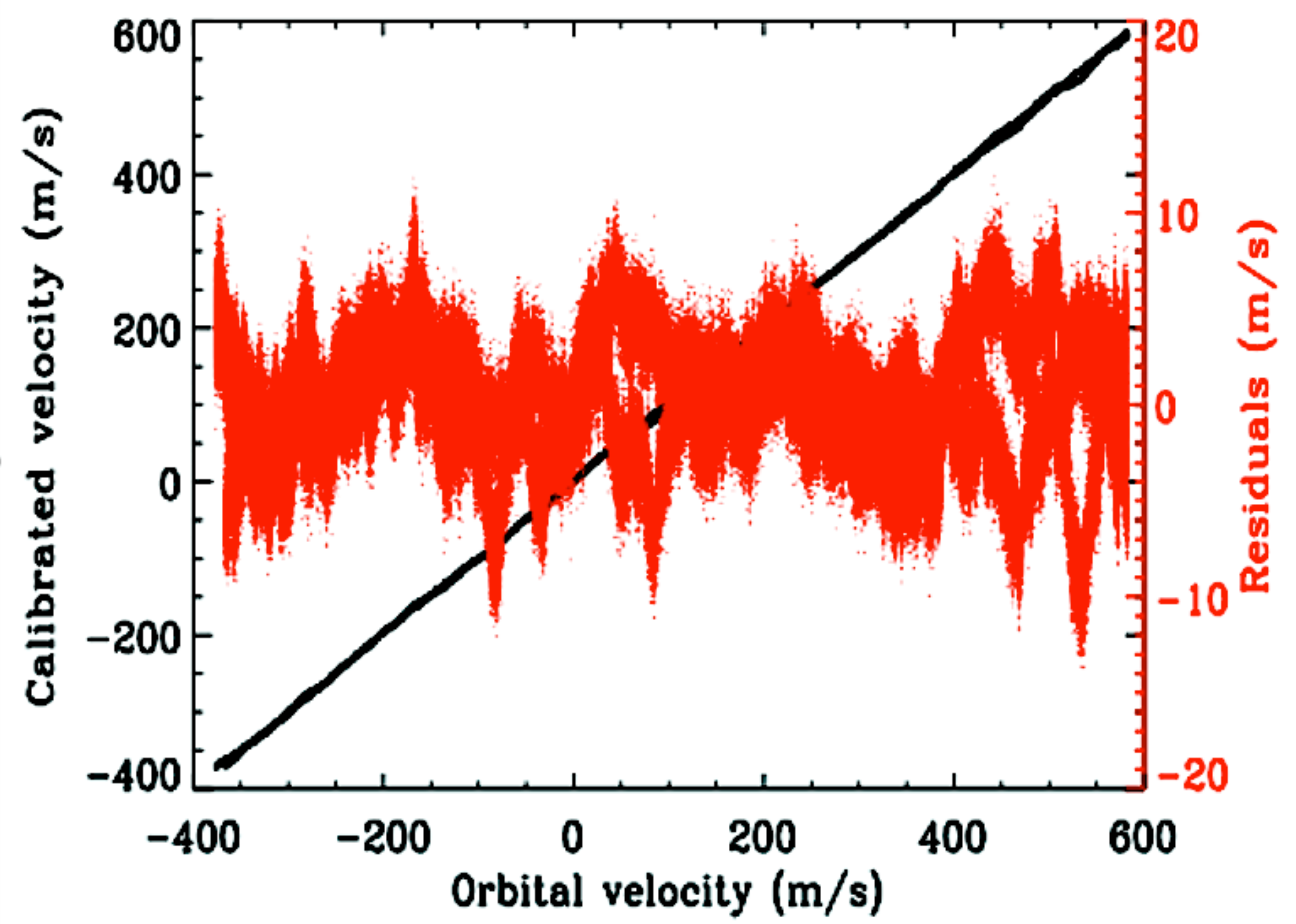}      
  \caption{Calibrated velocity as a function of Sun-instrument
    line-of-sight velocity.  Residual velocities shown in red with
    scale on the right ordinate.  The impact of solar surface activity
    in the residuals 
  can be clearly seen as a near 13 day period fluctuation.  The
  magnitude of the solar oscillation signal is a few $\rm m \; s^{-1}$.}
  \label{davies:fig2}
\end{figure}

\begin{figure}[ht!]
 \centering
 \includegraphics[width=0.8\textwidth,clip]{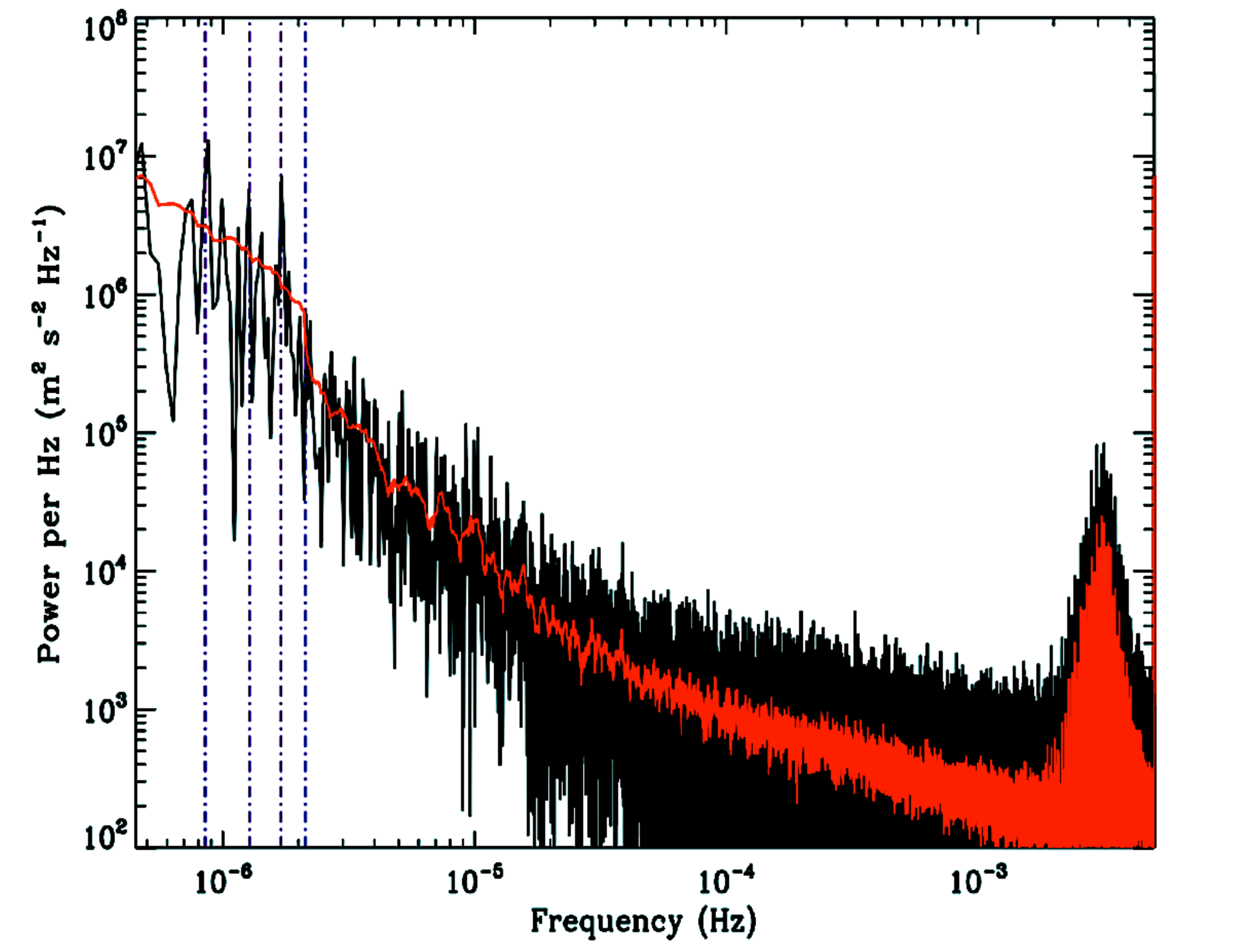}      
  \caption{Power spectrum of the 387 day residuals using the new
    correction and calibration method.  Black: power spectrum.  Red:
    20 point boxcar smoothed power spectrum.  Vertical lines:
    overtones of the Carrington rotation rate (highlighting peaks in
    the power spectrum due to solar surface activity).  The ``five
    minute'' modes are clearly visible at high frequency ($\approx 3 \;
    {\rm mHz}$).}
  \label{davies:fig3}
\end{figure}

We show an example of the new correction and calibration technique
applied to 387 days of GOLF data.  Figure \ref{davies:fig2} shows the strong
correspondence between the Sun-instrument line-of-sight velocity and
the calibrated velocities returned by the new method.  In addition,
Fig. \ref{davies:fig2} shows the residual velocities (measured velocity less $v_{\rm
  los}$) which show both the impact of solar surface activity (period
around 13 days) and solar oscillations (period of about 5 minutes and
amplitude of a few $\rm m \; s^{-1}$).  Figure \ref{davies:fig3} shows the power spectrum
of the residual velocities.  The frequency domain again shows the
presence of solar surface activity and solar oscillations. 

\section{Conclusions}

Here we present work focusing on achieving a stable correction and
calibration of GOLF data that could be applied to 15 years of
observations.  We demonstrate that we can achieve a good
correspondence between the Sun-instrument line-of-sight velocity and
the calibrated velocity.  The calibrated velocities show the
signatures of both solar surface activity and solar oscillations.

\begin{acknowledgements}
The GOLF instrument is a cooperative effort of many individuals, to whom we are indebted. SoHO is a mission of international cooperation between ESA and NASA. The authors thanks the support of the CNES.
\end{acknowledgements}

\bibliographystyle{aa}  
\bibliography{davies} 

\end{document}